\documentclass[twocolumn, prl, superscriptaddress,notitlepage]{revtex4-2}
\usepackage{graphicx}
\usepackage{dcolumn}
\usepackage{amsthm}  
\usepackage{amsmath} 
\usepackage{comment}
\usepackage{bm}
\usepackage[usenames,dvipsnames]{color}
\usepackage[most]{tcolorbox}
\usepackage{multirow}
\usepackage{gensymb}
\usepackage{amssymb}
\usepackage{amsmath}
\usepackage{xcolor}
\usepackage{algorithm}
\usepackage{algorithmic}
\usepackage[normalem]{ulem}
\usepackage{CJK}
\usepackage{comment}
\usepackage{amsfonts}
\usepackage[colorlinks, linkcolor=blue,anchorcolor=blue,citecolor=blue,urlcolor=blue]{hyperref}
\usepackage{amssymb}
\usepackage{pifont}
\usepackage{physics}
\usepackage{natbib}
\usepackage{xcolor}

\usepackage{color,soul}

\usetikzlibrary{arrows.meta,calc}

\definecolor{QECnavy}{HTML}{172239}
\definecolor{QECpurple}{HTML}{4E2A84}
\definecolor{QECteal}{HTML}{007F86}
\definecolor{QECmuted}{HTML}{596579}
\definecolor{QECred}{HTML}{B93D50}
\definecolor{QEClight}{HTML}{F3F0F7}

\tikzset{
  QECtext/.style={
    anchor=north west,
    align=left,
    inner sep=0pt,
    text=QECnavy,
    font=\fontsize{10}{12.7}\selectfont
  },
  QECsmall/.style={
    QECtext,
    font=\fontsize{8.8}{11}\selectfont
  },
  QECsub/.style={
    QECtext,
    font=\bfseries\fontsize{10.5}{13}\selectfont,
    text=QECpurple
  },
  QECarr/.style={
    -{Stealth[length=2.0mm,width=1.5mm]},
    line width=0.8pt,
    draw=QECmuted
  }
}

\begin{document}

\title{Optimal Error-corrected Hamiltonian estimation with nuisance parameters}

\author{Zhiyao Hu }\email[]{zhiyaohu.phys@gmail.com}

\affiliation{Chicago Quantum Institute, The University of Chicago, Illinois, USA}
\affiliation{Pritzker School of Molecular Engineering, The University of Chicago, Chicago, Illinois, USA}

\author{Haidong Yuan}\email[]{hdyuan@mae.cuhk.edu.hk}
\affiliation{
   Department of Mechanical and Automation Engineering, The Chinese University of Hong Kong, Shatin, Hong Kong}

\author{Liang Jiang}\email[]{liangjiang@uchicago.edu}

\affiliation{Chicago Quantum Institute, The University of Chicago, Illinois, USA}
\affiliation{Pritzker School of Molecular Engineering, The University of Chicago, Chicago, Illinois, USA}

\author{Zain H. Saleem}\email[]{zsaleem@anl.gov}
\affiliation{Mathematics and Computer Science Division, Argonne National Laboratory, Lemont, IL, USA}

\begin{abstract}
Realistic quantum sensing is often affected by both noise and unknown nuisance parameters. Under Markovian dynamics, existing error-correction approaches have often sought to suppress nuisance responses alongside environmental noise. Here we show that this local suppression of nuisance parameters is unnecessary and can be overly restrictive: nuisance parameters and Markovian noise impose fundamentally different constraints and require distinct treatments. We establish a general framework for quantum sensing with parameters encoded in a general time-dependent Hamiltonian, in the presence of both nuisance parameters and Markovian noise. We derive a necessary and sufficient condition for Heisenberg scaling, determine the exact optimal precision, and construct protocols that attain it. Our results show that to achieve Heisenberg scaling Markovian noise must be suppressed locally by quantum error correction, whereas nuisance parameters need not, and in general should not, be eliminated locally. Instead, optimally designed coherent control can separate their accumulated effects from the target signal over the full sensing evolution. Consequently, Heisenberg scaling can remain attainable even when the target signal is instantaneously aligned with nuisance-parameter directions, in sharp contrast to alignment with Markovian noise. This distinction substantially enlarges the regime in which Heisenberg scaling is attainable and provides a general principle for designing robust quantum sensors.
\end{abstract}

\maketitle

\paragraph{Introduction---}
Quantum metrology harnesses quantum mechanical effects such as superposition and entanglement to enhance sensing precision beyond the standard quantum limit (SQL), with applications ranging from spectroscopy and field sensing to imaging and fundamental tests of physics \cite{introhelstrom1969quantum,introholevo2001statistical,introholevo2011probabilistic,qmMaccone_2004_review,qmGiovannetti_2011_review,qmdegen2017quantum,qmtoth2014quantum,qmpezze2018quantum,polino2020photonic,demille2024quantum,liu2024fully,liu2022optimal,lawrie2019quantum,pirandola2018advances,fujiwara2008fibre,hou2021zero,hou2020minimal,isogawa2025entanglement,maze2008nanoscale,taylor2008high, cao2025joint,li2023improving,zaporski2025quantum}. For single-parameter sensing under time-independent unitary dynamics, the Heisenberg limit (HL) characterizes the ultimate attainable precision scaling \cite{escher2011general,demkowicz2012elusive}. In this ideal setting, the HL can be achieved through optimal probe preparation, coherent control, and measurement \cite{qmGiovannetti_2006_metrology, yuan2015optimal}.

Realistic sensing, however, must contend with both environmental noise and additional unknown parameters, known as nuisance parameters \cite{suzuki2020quantum}. 
For Markovian noise, quantum error correction (QEC) can protect the parameter encoding while suppressing environmental errors. In the time-independent setting, the Hamiltonian-not-in-Lindblad-span (HNLS) condition provides a necessary and sufficient criterion for retaining Heisenberg scaling \cite{zhou2018achieving, demkowicz2017adaptive}. Related approaches have subsequently been developed for time-dependent noisy dynamics \cite{previdi2026precision} and for sensing in the presence of nuisance parameters \cite{abbasgholinejad2026error}.

The interplay between Markovian noise and nuisance parameters becomes particularly intricate under time-dependent dynamics, as noise correction and nuisance-robust estimation impose distinct requirements on the same sensing protocol. QEC must suppress environmental errors while preserving the target encoding \cite{zhou2018achieving,zhou2021asymptotic,demkowicz2017adaptive}, whereas nuisance-robust estimation further requires the retained target response to remain distinguishable from those of other unknown parameters \cite{hu2026optimal}. Since the choice of code and control affects both noise suppression and parameter distinguishability, it is unclear how much information about the target can be preserved while satisfying both requirements. Consequently, the fundamental precision limits and the sensing protocols that attain them remain to be established.

In this Letter, we develop a general framework for estimating parameters encoded in a general time-dependent Hamiltonian in the simultaneous presence of unknown nuisance parameters and Markovian noise. We identify the coherent HL contribution in general time-dependent dynamics, and derive a necessary and sufficient condition for achieving it and determine its exact optimal value. We further construct an explicit sensing protocol combining probe preparation, coherent control, QEC, and measurement that attains this optimum.
Our framework reveals a fundamental distinction between the two sources of uncertainty: Markovian noise must be suppressed locally by QEC, whereas nuisance responses can be retained locally and coherently separated from the target over the full sensing evolution. Consequently, the HL can remain attainable even when the target signal is instantaneously aligned with nuisance directions, a regime that would be excluded by treating nuisance parameters on the same footing as Markovian noise. 

\paragraph{Settings---}
We consider a vector of unknown parameters
$\boldsymbol{x}=(x_1,\ldots,x_N)^T$ encoded in a general time-dependent Hamiltonian, while the probe is simultaneously subject to Markovian noise. The dynamics are described by
\begin{equation}
\dot{\rho}_{\boldsymbol{x}}
=
-i[H_{\rm tot}(\boldsymbol{x},t),\rho_{\boldsymbol{x}}]
+
\sum_j \mathcal{D}[L_j(t)]\rho_{\boldsymbol{x}},
\label{eq:lindblad}
\end{equation}
where $H_{\rm tot}(\boldsymbol{x},t)=H(\boldsymbol{x},t)+H_C(t)$, with $H_C(t)$ denoting the control Hamiltonian, and
$\mathcal{D}[L]\rho=L\rho L^\dagger-\frac{1}{2}\{L^\dagger L,\rho\}$.
Both $H_C(t)$ and the Lindblad operators $L_k(t)$ are independent of the unknown parameters.

Our objective is to estimate $x_1$ while the remaining unknown parameters $\boldsymbol{x}_{\mathrm n}=(x_2,\ldots,x_N)^T$ act as nuisance parameters. Their presence reduces the information that can be unambiguously attributed to $x_1$. For $\nu$ independent repetitions, the quantum Cramér--Rao bound (QCRB) gives \cite{qcrbfisher1925theory,qcrbrao1992information,introhelstrom1969quantum,qcrbrao1992information,qmCaves_1994_Bures}
\begin{equation}
\operatorname{Var}(\hat{x}_1)
\ge
\frac{[\boldsymbol J_{\boldsymbol{x}}^{-1}]_{11}}{\nu}=
\frac{1}{\nu J_{x_1|\boldsymbol{x}_{\mathrm n}}},
\label{eq:nuisance_qcrb}
\end{equation}
where $\boldsymbol J_{\boldsymbol{x}}$ is the quantum Fisher information matrix (QFIM) and $J_
{x_1|\boldsymbol{x}_{\mathrm n}}
=\frac{1}{[\boldsymbol J_{\boldsymbol{x}}^{-1}]_{11}}$ denotes the effective quantum Fisher information (QFI) for estimating $x_1$ in the presence of the nuisance parameters. For a nonsingular QFIM, this effective QFI admits the equivalent variational representation \cite{liu2020quantum,hu2026optimal}
\begin{equation}
J_{x_1|\boldsymbol{x}_{\mathrm n}}
=
\min_{\boldsymbol{\lambda}\in\mathbb R^{N-1}}
\begin{pmatrix}
1 & -\boldsymbol{\lambda}^{T}
\end{pmatrix}
\boldsymbol J_{\boldsymbol{x}}
\begin{pmatrix}
1\\
-\boldsymbol{\lambda}
\end{pmatrix},
\label{eq:nuisance_variational}
\end{equation}
where $\boldsymbol{\lambda}=(\lambda_2,\ldots,\lambda_N)^T$. 
For each fixed $\boldsymbol{\lambda}$, the quadratic form is the QFI $J_{\tilde{x}_1}$ of a local one-parameter submodel parameterized by $\tilde{x}_1$ whose tangent direction is given by $\partial_{\boldsymbol{\lambda}} = \partial_{x_1} - \sum_{j=2}^{N}\lambda_j\partial_{x_j}.$ Equivalently, the parameter $\tilde{x}_1$ satisfies $ \frac{d\boldsymbol{x}}{d\tilde{x}_1} = \begin{pmatrix} 1\\ -\boldsymbol{\lambda} \end{pmatrix}, $ so that its QFI is  $J_{\tilde{x}_1} = \begin{pmatrix} 1 & -\boldsymbol{\lambda}^{T} \end{pmatrix} \boldsymbol J_{\boldsymbol{x}} \begin{pmatrix} 1\\ -\boldsymbol{\lambda} \end{pmatrix}.$ This equivalence allows the use of single-parameter channel bounds to each effective parameter $\tilde{x}_1$. 



For the time-dependent Markovian dynamics in Eq.~\eqref{eq:lindblad}, the infinitesimal channel from $t$ to $t+dt$ admits the Kraus representation \begin{equation} \mathcal E^{\boldsymbol{x}}_{t+dt,t}(\rho) = \sum_{j=0}^{r}K_j(\boldsymbol{x},t,dt)\rho K_j^\dagger(\boldsymbol{x},t,dt) +O(dt^2), \end{equation} with \begin{align} K_0(\boldsymbol{x},t,dt) &= I-\left[ iH_{\rm tot}(\boldsymbol{x},t) +\frac{1}{2}\sum_{j=1}^{r}L_j^\dagger(t)L_j(t) \right]dt, \nonumber\\ K_j(\boldsymbol{x},t,dt) &= \sqrt{dt}\,L_j(t), \qquad j=1,\ldots,r . \label{eq:kraus_expansion} 
\end{align} 
The Kraus representation is not unique: a parameter-dependent unitary rotation, $u(\boldsymbol{x},t)$ on the Kraus operators, $\widetilde K_l=\sum_j u_{lj}(\boldsymbol{x},t)K_j$, leaves the physical channel unchanged. Applying the continuous-time channel-extension bound and subsequently minimizing over $\boldsymbol{\lambda}$ gives \cite{zhou2018achieving,zhou2021asymptotic,previdi2026precision}
\begin{equation} 
\sup_{\mathcal P} J_{x_1|\boldsymbol{x}_{\mathrm n}}^{\mathcal P}(T) \le 4\min_{\boldsymbol{\lambda},h(t)} \left[ \sqrt{A_{\boldsymbol{\lambda}}^{h}(T)} + B_{\boldsymbol{\lambda}}^{h}(T) \right]^2, \label{eq:nuisance_ce_bound} 
\end{equation} 
here
\(A_{\boldsymbol{\lambda}}^{h}(T) = \int_0^T \left\| \alpha_{\boldsymbol{\lambda}}^{h}(t) \right\|dt,\) 
 \(B_{\boldsymbol{\lambda}}^{h}(T) = \int_0^T \left\| \beta_{\boldsymbol{\lambda}}^{h}(t) \right\|dt,\) 
with 
\begin{align} 
\alpha_{\boldsymbol{\lambda}}^{h}(t) &= \lim_{dt\to0}\frac{1}{dt} \sum_l \left( \partial_{\boldsymbol{\lambda}}\widetilde K_l(\boldsymbol x,t) \right)^\dagger \left( \partial_{\boldsymbol{\lambda}}\widetilde K_l(\boldsymbol x,t) \right), \nonumber\\ \beta_{\boldsymbol{\lambda}}^{h}(t) &= \lim_{dt\to0}\frac{i}{dt} \sum_l \left( \partial_{\boldsymbol{\lambda}}\widetilde K_l(\boldsymbol x,t) \right)^\dagger \widetilde K_l(\boldsymbol x,t), \label{eq:alpha_beta} 
\end{align} 
here $\partial_{\boldsymbol{\lambda}}\widetilde K_l=\sum_{jk}u_{lj}\partial_{\boldsymbol{\lambda}}K_j-iu_{lk} h_{kj}K_j$ with $h(t)=iu^\dagger(t)\partial_{\boldsymbol{\lambda}}u(t)$ representing the gauge freedom of the Kraus operators at each time point ~\cite{zhou2018achieving}, $T$ is the interrogation time, $\|\cdot\|$ denotes the operator norm. The optimization is over all admissible sequential sensing protocols $\mathcal P$, which includes the preparation of any initial probe state and the implementation of any intermediate control during the evolution. The detailed derivation of Eq.~\eqref{eq:nuisance_ce_bound}
can be found in the Supplemental Material.

The two terms in Eq.~\eqref{eq:nuisance_ce_bound} characterize distinct contributions to the attainable QFI. For time-independent dynamics, the $B_{\boldsymbol{\lambda}}^h$ term gives rise to the conventional Heisenberg-limited (HL) scaling $T^2$. Motivated by this structure, we denote the HL contribution for general time-dependent dynamics as
\begin{equation}
J_{x_1|\boldsymbol{x}_{\mathrm n}}^{\mathrm{HL}}(T)
:=
4\min_{\boldsymbol{\lambda},h(t)}
\left[
B_{\boldsymbol{\lambda}}^{h}(T)
\right]^2 .
\label{eq:HL_channel_bound}
\end{equation}
We now determine when Heisenberg scaling is attainable and its exact optimal value in the simultaneous presence of Markovian noise and nuisance parameters. Our analysis reveals that the two impose fundamentally different temporal constraints: Markovian noise should be suppressed locally in time, whereas nuisance responses need not be eliminated instantaneously and only need to be separated from the target over the full sensing evolution.

\begin{figure}[htbp]
\includegraphics[width=0.45\textwidth]{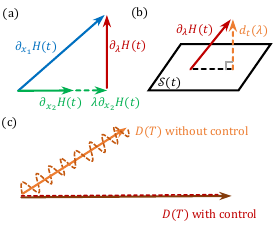}
\caption{
(a) Subtracting the nuisance response
$\lambda\partial_{x_2}H(t)$ from the target response
$\partial_{x_1}H(t)$ defines the effective Hamiltonian response
$\partial_\lambda H(t)=\partial_{x_1}H(t)
-\lambda\partial_{x_2}H(t)$, with $\lambda$ fixed throughout
the evolution.
(b) The INLTR $d_t(\lambda)$ is the spectral-width distance
from the effective Hamiltonian response $\partial_\lambda H(t)$
to the instantaneous Hermitian Lindblad span $\mathfrak S(t)$.
It quantifies the component outside this span that contributes
to HL scaling after exact QEC.
(c) Without control, contributions to the logical generators
at different times are generally misaligned and partially
cancel. With the optimal QEC code, control aligns the retained
effective responses along a fixed logical direction,
allowing them to accumulate coherently and attain the
GNLTR $D(T)$.
Orange (red) dashed curves trace the instantaneous responses
without (with) control, while solid arrows represent the
corresponding logical generators.
}
\label{ci}
\end{figure}

\paragraph{Main results---}
The optimizations over $h(t)$ and $\boldsymbol{\lambda}$ in Eq.~\eqref{eq:HL_channel_bound} have fundamentally different temporal structures. For any fixed $\boldsymbol{\lambda}$, the Kraus gauge $h(t)$ can be optimized independently at each time point so that, under standard regularity conditions, 
\begin{equation} 
\min_{h(\cdot)} \int_0^T \left\|\beta_{\boldsymbol{\lambda}}^h(t)\right\|dt = \int_0^T \min_{h(t)} \left\|\beta_{\boldsymbol{\lambda}}^h(t)\right\|dt . \label{eq:pointwise_gauge} 
\end{equation}
The pointwise minimum has the geometric form, \(\min_{h(t)} \left\| \beta_{\boldsymbol{\lambda}}^{h}(t) \right\|= \frac{1}{2}d_t(\boldsymbol{\lambda})\), where 
\begin{equation} 
d_t(\boldsymbol{\lambda}) = \min_{G(t)\in\mathfrak S(t)} \left|\partial_{\boldsymbol{\lambda}}H(\boldsymbol{x},t)-G(t) \right|_s,
\label{eq:instantaneous_distance} 
\end{equation} 
here $|\cdot|_s$ denotes the spectral width and 
\begin{equation} 
\mathfrak S(t) = \operatorname{span}_{\mathbb R} \left\{ I,L_k(t),L_k^\dagger(t),L_j^\dagger(t)L_k(t) \right\}_{j,k}^{\mathrm{Herm}} 
\end{equation}
is the instantaneous Hermitian Lindblad span \cite{zhou2018achieving, demkowicz2017adaptive}.
We refer to $d_t(\boldsymbol{\lambda})$ as the instantaneous nuisance-aware Lindblad-transverse response (INLTR). It quantifies the component of the effective Hamiltonian response that remains outside the instantaneous Lindblad span and is therefore contributing to HL scaling after exact correction of the Markovian errors. The pointwise optimization in Eq.~\eqref{eq:pointwise_gauge} reflects the local nature of this requirement: the QEC code can be adapted at each time point to suppress the instantaneous Markovian errors while retaining the maximal logical response.

The treatment of nuisance parameters is fundamentally different as $\lambda$ is a constant vector instead of a function of time. Substituting Eq.~\eqref{eq:instantaneous_distance} into Eq.~\eqref{eq:HL_channel_bound} gives
\begin{equation}
J_{x_1|\boldsymbol{x}_{\mathrm n}}^{\mathrm{HL}}(T)
=
\min_{\boldsymbol{\lambda}\in\mathbb R^{N-1}}
\left[\int_0^T d_t(\boldsymbol{\lambda})\,dt\right]^2
=D^2(T),
\label{eq:optimal_HL_precision}
\end{equation}
where $D(T):=\min_{\boldsymbol{\lambda}\in\mathbb R^{N-1}}\int_0^T d_t(\boldsymbol{\lambda})\,dt$ is the global nuisance-aware Lindblad-transverse response (GNLTR). Consequently, a nonzero nuisance-robust HL contribution exists if and only if
\begin{equation}
D(T)>0.
\label{eq:HL_condition}
\end{equation}
Here the minimization over $\boldsymbol{\lambda}$ remains outside the time integral: a single $\boldsymbol{\lambda}$ characterizes the nuisance ambiguity over the entire sensing evolution. Unlike Markovian errors, nuisance responses need not be suppressed at every instant. They may overlap with the target response locally, provided that their accumulated effects can be separated from the target after the full evolution.  
This local--global distinction is encoded directly in the nested optimization
\begin{equation}
D(T)=\underbrace{\min_{\boldsymbol{\lambda}}}_{\mathrm{global\ nuisance}}
\int_0^T dt\,
\underbrace{\min_{G(t)\in\mathfrak S(t)}}_{\mathrm{local\ error}}
\left|
\partial_{\boldsymbol{\lambda}}H(\boldsymbol{x},t)-G(t)
\right|_s .
\end{equation}
The inner minimization is performed independently at each instant and reflects the local QEC constraint imposed by Markovian noise, whereas the outer minimization acts on the accumulated response and therefore imposes only a global constraint on the nuisance parameters.

This distinction becomes particularly transparent in the time-independent limit. When $H(\boldsymbol{x},t)=H(\boldsymbol{x})$ and $L_j(t)=L_j$, both $\mathfrak S$ and $d(\lambda)=\min_{G\in\mathfrak S}
\left|
\partial_{\boldsymbol{\lambda}}H(\boldsymbol{x})-G
\right|_s$ are time independent. Equation~\eqref{eq:optimal_HL_precision} then reduces to 
\begin{align} 
J_{x_1|\boldsymbol{x}_{\mathrm n}}^{\mathrm{HL}}(T) &= T^2 \left[ \min_{\boldsymbol{\lambda},\,G\in\mathfrak S} \left| \partial_{\boldsymbol{\lambda}}H(\boldsymbol{x})-G \right|_s \right]^2 \nonumber\\ &= T^2 \left[ \min_{G\in\mathfrak S_{\mathrm n}} \left| \partial_{x_1}H(\boldsymbol{x})-G \right|_s \right]^2 , \label{eq:time_independent_HL} 
\end{align} 
where 
\begin{equation}
\mathfrak S_{\mathrm n}
=
\mathfrak S
+
\operatorname{span}_{\mathbb R}
\left\{
\partial_{x_2}H,\ldots,\partial_{x_N}H
\right\}.
\label{eq:nuisance_lindblad_span}
\end{equation}
is the Lindblad span enlarged by the nuisance-parameter directions.
Thus, in the time-independent setting, nuisance directions can be incorporated into a single static enlarged span, and the condition Eq.~\eqref{eq:HL_condition} reduces to
$\partial_{x_1}H\notin\mathfrak S_{\mathrm n}$,
recovering the generalized HNLS condition in the presence of nuisance parameters \cite{abbasgholinejad2026multiparameter}. 

For time-dependent dynamics, however, this reduction no longer holds. Including the instantaneous nuisance directions in an enlarged Lindblad span at each time would effectively allow their coefficients $\boldsymbol{\lambda}$ to vary independently with $t$. This replaces the global minimization 
\(
\min_{\boldsymbol{\lambda}} \int_0^T d_t(\boldsymbol{\lambda})\,dt \)
by the more restrictive local minimization 
\(
\int_0^T \min_{\boldsymbol{\lambda}} d_t(\boldsymbol{\lambda})\,dt . \)
The latter requires nuisance responses to be removed instantaneously, whereas nuisance-robust estimation requires only their accumulated effects to be separated from the target \cite{hu2026optimal}. Thus, a target response may be instantaneously indistinguishable from nuisance directions and yet remain distinguishable over the full evolution. Locally treating nuisance parameters on the same footing as Markovian noise can discard useful target information and impose unnecessarily restrictive conditions for achieving the HL.

The GNLTR can be evaluated efficiently because the optimization defining $D(T)$ is convex. Introducing scalar functions $a(t)$ and $b(t)$ that bound the extremal eigenvalues of the Lindblad-transverse response, we obtain the semidefinite program (SDP)
\begin{equation}
\begin{aligned}
\min_{\boldsymbol{\lambda},G,a,b}\quad
&\int_0^T [a(t)-b(t)]\,dt,\\
\mathrm{s.t.}\quad
&b(t)I\preceq
\partial_{\boldsymbol{\lambda}}H(\boldsymbol{x},t)-G(t)
\preceq a(t)I,\\
&G(t)\in\mathfrak S(t).
\end{aligned}
\label{eq:optimal_lambda_sdp}
\end{equation}
At the optimum, $a(t)$ and $b(t)$ are respectively the largest
and smallest eigenvalues of
$\partial_\lambda H(\boldsymbol{x},t)-G(t)$, yielding $a(t)-b(t)
=
\left|\partial_\lambda H(\boldsymbol{x},t)-G(t)\right|_s$.
For numerical implementation, discretizing $[0,T]$ into $M$ intervals with durations $\Delta t_k$ and representative times $t_k$ gives
\begin{equation}
\begin{aligned}
\min_{\boldsymbol{\lambda},\{G_k,a_k,b_k\}}\quad
&\sum_{k=1}^M\Delta t_k(a_k-b_k),\\
\mathrm{s.t.}\quad
&b_kI\preceq
\partial_{\boldsymbol{\lambda}}H(\boldsymbol{x},t_k)-G_k
\preceq a_kI,\\
&G_k\in\mathfrak S(t_k).
\end{aligned}
\label{eq:discretized_lambda_sdp}
\end{equation}
Under standard regularity assumptions, the optimal value converges to that of the continuous-time problem as the discretization is refined.

We next show that the optimal HL contribution is attainable and that the attaining construction realizes precisely this local--global distinction.
Let $\boldsymbol{\lambda}^\star$ be a minimizer
of Eq.~\eqref{eq:optimal_lambda_sdp}.
The INLTR admits the dual representation
\begin{equation}
d_t(\boldsymbol{\lambda})
=
\max_{\rho_0,\rho_1}
\operatorname{Tr}\left[
(\rho_0-\rho_1)
\partial_{\boldsymbol{\lambda}}H(\boldsymbol{x},t)
\right],
\label{eq:dual_distance}
\end{equation}
where the maximization is over density operators
satisfying
\begin{equation}
\operatorname{Tr}[(\rho_0-\rho_1)G]=0,
\qquad
\forall G\in\mathfrak S(t).
\label{eq:dual_qec_condition}
\end{equation}
Choose the optimal $\rho_0(t)$ and $\rho_1(t)$
for $\boldsymbol{\lambda}^\star$ at each $t$.
Let $\ket{\psi_j(t)}$ be purifications of $\rho_j(t)$,
and introduce orthogonal flag states to define the logical codewords
\begin{equation}
\ket{0_L(t)}
=
\ket{\psi_0(t)}\ket{0}_F,
\qquad
\ket{1_L(t)}
=
\ket{\psi_1(t)}\ket{1}_F.
\end{equation}
For
$P(t)=\ket{0_L(t)}\!\bra{0_L(t)}+\ket{1_L(t)}\!\bra{1_L(t)}$,
Eq.~\eqref{eq:dual_qec_condition} and the orthogonality of the flags imply
\begin{equation}
P(t)GP(t)
=
c_G(t)P(t),
\qquad
\forall G\in\mathfrak S(t),
\label{eq:exact_qec_condition}
\end{equation}
where $c_G(t)$ is independent of the logical state.
Thus, at every instant, the logical code satisfies the exact QEC condition for the Markovian errors.

Importantly, exact correction of the noise does not require the parameter responses to vanish. Within the same logical subspace, 
\begin{equation}
P(t)\partial_{x_j}H(\boldsymbol{x},t)P(t)
=
{\frac{1}{2}}a_j(t)P(t)
+
\frac{r_j(t)}{2}Z_L(t),
\label{eq:projected_parameter_velocity}
\end{equation}
where
\begin{equation}
r_j(t)
=
\operatorname{Tr}\left[
(\rho_0(t)-\rho_1(t))
\partial_{x_j}H(\boldsymbol{x},t)
\right],
\label{eq:parameter_responses}
\end{equation}
$a_j(t)=\operatorname{Tr}
[(\rho_0(t)+\rho_1(t))\partial_{x_j}H(\boldsymbol{x},t)]$,
and 
$Z_L(t)=\ket{0_L(t)}\!\bra{0_L(t)}
-\ket{1_L(t)}\!\bra{1_L(t)}$.
Hence, $r_j(t)$ is the instantaneous relative-phase response to $x_j$
retained after correcting the Markovian noise.

To accumulate these retained responses, we prepare the balanced logical probe
\begin{equation}
\ket{\varphi_0^\star}
=
\frac{1}{\sqrt{2}}
\left(
\ket{0_L(0)}
+
\ket{1_L(0)}
\right)
\label{eq:optimal_logical_probe}
\end{equation}
and use coherent control to transport the logical basis along the time-dependent code,
\begin{equation}
U_{\rm opt}(t)\ket{j_L(0)}
=
\ket{j_L(t)},
\qquad j=0,1.
\label{eq:optimal_logical_control}
\end{equation}
At a local operating point $\boldsymbol{x}_0$,
this transport can be achieved with the coherent control
\begin{equation}
H_C(t)
=
i\dot U_{\rm opt}(t)U_{\rm opt}^\dagger(t)
-
H(\boldsymbol{x}_0,t).
\label{eq:optimal_control_hamiltonian}
\end{equation}

Under ideal continuous QEC, the resulting logical evolution is unitary and the parameter generators are~\cite{qmpang_2014_metrology,qmpang_optimal_2017}
\begin{align}
S_{x_j}^{L}(T)
&=
\int_0^T
U_{\rm opt}^\dagger(t)
P(t)\partial_{x_j}H(\boldsymbol{x}_0,t)P(t)
U_{\rm opt}(t)\,dt
\nonumber\\
&=
c_j I_L
+
\frac12
\left[\int_0^T r_j(t)\,dt\right]Z_L(0),
\end{align}
where $c_j=\frac{1}{2}\int_0^T a_j(t)\,dt$ and we have used
$U_{\rm opt}^\dagger(t)Z_L(t)U_{\rm opt}(t)=Z_L(0)$.
Substituting
$\partial_{\boldsymbol{\lambda}}H
=\partial_{x_1}H-\sum_{j=2}^N\lambda_j\partial_{x_j}H$
into Eq.~\eqref{eq:dual_distance} and using
Eq.~\eqref{eq:parameter_responses}, we obtain
\begin{equation}
d_t(\boldsymbol{\lambda}^\star)
=
r_1(t)-\sum_{j=2}^N\lambda_j^\star r_j(t).
\label{eq:dual_response}
\end{equation}

As shown in the Supplemental Material, the dual-optimal states
at $\boldsymbol{\lambda}^\star$ can be chosen so that all
accumulated nuisance responses vanish,
\begin{equation}
\int_0^T r_j(t)\,dt=0,
\qquad j=2,\ldots,N.
\label{eq:vanishing_nuisance_response}
\end{equation}
Integrating Eq.~\eqref{eq:dual_response} then gives
\begin{equation}
\int_0^T r_1(t)\,dt
=
\int_0^T d_t(\boldsymbol{\lambda}^\star)\,dt
=
D(T).
\label{eq:optimal_target_response}
\end{equation}
The optimal logical generators therefore reduce to
\begin{align}
S_{x_1}^{L}(T)
&=
c_1 I_L
+
\frac12
\left[
\int_0^T d_t(\boldsymbol{\lambda}^\star)\,dt
\right]Z_L(0),
\nonumber\\
S_{x_\ell}^{L}(T)
&=
c_\ell I_L,
\qquad \ell=2,\ldots,N.
\label{eq:optimal_logical_generators}
\end{align}



Since the nuisance generators in Eq. \eqref{eq:optimal_logical_generators} are proportional to the logical identity, they carry no relative-phase information in the final logical state. The effective QFI of the pure logical probe reduces to \(J_{x_1|\boldsymbol{x}_{\mathrm n}}^{\mathrm{eff}}(T)=4\operatorname{Var}(S^{L}_{x_1})\) \cite{qmpang_optimal_2017,hu2024control,hu2025optimal,hu2026optimal}. The balanced logical probe maximizes this variance, yielding
\begin{equation}
J_{x_1|\boldsymbol{x}_{\mathrm n}}^{\mathrm{eff}}(T)
=
\left[
\int_0^T
d_t(\boldsymbol{\lambda}^\star)\,dt
\right]^2
=D^2(T)=
J_{x_1|\boldsymbol{x}_{\mathrm n}}^{\mathrm{HL}}(T),
\label{eq:qec_attained_HL_precision}
\end{equation}
which establishes attainability.

Finally, necessity follows directly from the GNLTR condition. If $D(T)=0$, there exists $\boldsymbol{\lambda}_0$ such that
$\int_0^T d_t(\boldsymbol{\lambda}_0)\,dt=0$.
Since $d_t(\boldsymbol{\lambda}_0)\ge0$, this implies $d_t(\boldsymbol{\lambda}_0)=0$ almost everywhere and hence
$\partial_{\boldsymbol{\lambda}_0}H(\boldsymbol{x},t)\in\mathfrak S(t)$
almost everywhere along the evolution. 
The effective signal direction $\partial_{\boldsymbol{\lambda}_0}H$ is then locally indistinguishable from the Lindblad span and cannot retain an HL contribution under exact QEC.
Equation~\eqref{eq:optimal_HL_precision} then gives
$J_{x_1|\boldsymbol{x}_{\mathrm n}}^{\mathrm{HL}}(T)=0$,
establishing necessity and completing the proof.


The same framework applies directly to functional sensing, where the quantity of interest is a prescribed linear function $\theta_1=\boldsymbol{c}^T\boldsymbol{x}$ of multiple unknown parameters $\boldsymbol{x}=(x_1,\ldots,x_N)^T$ encoded in the Hamiltonian $H(\boldsymbol{x},t)$, with $\boldsymbol{c}\neq\boldsymbol{0}$. To apply our framework, we introduce an invertible linear reparametrization $\boldsymbol{\theta}=A\boldsymbol{x}=(\theta_1,\boldsymbol{\theta}_{\mathrm n}^T)^T$, where the first row of $A$ is $\boldsymbol{c}^T$ and the remaining rows define the nuisance parameters. Since $\boldsymbol{x}=A^{-1}\boldsymbol{\theta}$, varying $\theta_1$ while holding $\boldsymbol{\theta}_{\mathrm n}$ fixed defines the direction 
\begin{equation} 
\boldsymbol v = A^{-1}\boldsymbol e_1, \qquad \partial_{\theta_1} = \sum_{j=1}^{N}v_j\partial_{x_j}, \label{eq:functional_direction} \end{equation} 
where $\boldsymbol e_1=(1,0,\ldots,0)^T$. Because the first row of $A$ is $\boldsymbol c^T$, this direction satisfies $\boldsymbol c^T\boldsymbol v=1$.
Different choices of the nuisance coordinates, i.e., of the remaining rows of $A$, generate different vectors $\boldsymbol v=A^{-1}\boldsymbol e_1$, and the set of all such directions is characterized by $\boldsymbol c^T\boldsymbol v=1$. Applying our main result and minimizing over this freedom gives
\begin{equation}
J_{\theta_1|\boldsymbol{\theta}_{\mathrm n}}^{\mathrm{HL}}(T)
=
\left[
\min_{\boldsymbol{c}^T\boldsymbol{v}=1}
\int_0^T
\min_{G(t)\in\mathfrak S(t)}
\left|
\sum_{j=1}^N v_j\partial_{x_j}H(\boldsymbol{x},t)-G(t)
\right|_s dt
\right]^2.
\label{eq:functional_HL}
\end{equation}
The resulting expression depends only on the target functional $\boldsymbol c^T\boldsymbol x$ and is independent of the particular choice of nuisance coordinates. Its optimal value is attainable by the same time-dependent QEC and coherent-control construction developed above. A detailed derivation is provided in the Supplemental Material.

\emph{Summary---}
We have established a general framework for optimal quantum sensing with parameters encoded in a time-dependent Hamiltonian in the simultaneous presence of Markovian noise and unknown nuisance parameters. We derived a necessary and sufficient condition for a nonzero HL contribution, determined its exact optimal value, and constructed a sensing protocol that attains it. Our results reveal that Markovian noise and nuisance parameters impose fundamentally different temporal constraints: environmental errors require local correction, whereas nuisance responses need not be suppressed instantaneously and only need to be separated from the target after accumulation over the full sensing evolution. Consequently, target information can retain Heisenberg scaling even when the target and nuisance responses overlap instantaneously. This distinction identifies a broader regime of Heisenberg-limited sensing than obtained by treating nuisance parameters as additional errors to be removed locally, and provides a general principle for jointly designing QEC and coherent control in realistic quantum sensing.

We acknowledge support from the Quantum Science and Technology-National Science and Technology Major Project (2023ZD0300600), the Research Grants Council of Hong Kong (14309223, 14309624, 14309022), Guangdong Provincial Quantum Science Strategic Initiative (GDZX2303007,GDZX2505003) and 1+1+1 CUHK-CUHK(SZ)-GDST Joint Collaboration Fund (Grant No. GRDP2025-022), ARO(W911NF-23-1-0077), ARO MURI (W911NF-21-1-0325), AFOSR MURI (FA9550-21-1-0209, FA9550-23-1-0338), ONR MURI (N000142612102), DARPA (HR0011-24-9-0361), NSF (OSI-2326767, OSI-2426975, EEC-2550064, OSI-2553578, OSI-2553619). This material is based upon work supported by the U.S. Department of Energy, Office of Science, National Quantum Information Science Research Centers and Advanced Scientific Computing Research (ASCR) program under contract number DE-AC02-06CH11357 as part of the InterQnet quantum networking project. This work was supported by Quantum Information Science Enabled Discovery 2.0 (QuantISED 2.0) for High Energy Physics (DE-FOA-0003354), with work at Argonne National Laboratory supported by the U.S. Department of Energy under Contract No. DE-AC02-06CH11357.

\framebox{\parbox{\linewidth}{
The submitted manuscript has been created by UChicago Argonne, LLC, Operator of 
Argonne National Laboratory (``Argonne''). Argonne, a U.S.\ Department of 
Energy Office of Science laboratory, is operated under Contract No.\ 
DE-AC02-06CH11357. 
The U.S.\ Government retains for itself, and others acting on its behalf, a 
paid-up nonexclusive, irrevocable worldwide license in said article to 
reproduce, prepare derivative works, distribute copies to the public, and 
perform publicly and display publicly, by or on behalf of the Government.  The 
Department of Energy will provide public access to these results of federally 
sponsored research in accordance with the DOE Public Access Plan. 
http://energy.gov/downloads/doe-public-access-plan.}}

\bibliography{main_text} 

\clearpage
\section*{End Matter}
We now illustrate the result using a qubit sensor probing a rotating
field in the presence of bit-flip noise,
\begin{equation}
\begin{aligned}
H(t)&=B\bigl[\cos(\omega t+\phi)\sigma_x
             +\sin(\omega t+\phi)\sigma_z\bigr],\\
L&=\sqrt{\gamma}\,\sigma_x,
\end{aligned}
\label{eq:frequency_example}
\end{equation}
where $B>0$ is known, $\omega$ is the target frequency, $\phi$ is
an unknown nuisance phase, and $\gamma>0$ is the noise strength.
At the operating point $(\omega_0,\phi_0)$, 
let $\vartheta_0(t)=\omega_0t+\phi_0$. 
The parameter derivatives are
\begin{equation}
\begin{aligned}
\partial_\omega H(t)
 &=Bt\bigl[-\sin\vartheta_0(t)\sigma_x
            +\cos\vartheta_0(t)\sigma_z\bigr],\\
\partial_\phi H(t)
 &=B\bigl[-\sin\vartheta_0(t)\sigma_x
           +\cos\vartheta_0(t)\sigma_z\bigr].
\end{aligned}
\label{eq:frequency_derivatives}
\end{equation}
Thus, \begin{equation} \partial_\omega H(t)=t\,\partial_\phi H(t). \label{eq:parallel_responses} \end{equation} The frequency and phase responses are therefore parallel at every instant. This example is particularly revealing: any strategy that removes the nuisance response locally also removes the target response, yet their different time dependence allows them to be separated over the full evolution.

For bit-flip noise,
$\mathfrak S=\operatorname{span}_{\mathbb R}\{I,\sigma_x\}$.
Along the effective direction $\partial_\lambda=\partial_\omega-\lambda\partial_\phi$, the INLTR is
\begin{equation}
\begin{aligned}
d_t(\lambda)
 &=\min_{G\in\mathfrak S}|\partial_\lambda H(t)-G|_s\\
 &=2B|t-\lambda|\,|\cos\vartheta_0(t)|.
\end{aligned}
\label{eq:frequency_distance}
\end{equation}
The GNLTR is therefore
\begin{equation}
\begin{aligned}
D(T)&=\min_{\lambda\in\mathbb R}\int_0^T d_t(\lambda)\,dt\\
 &=2B\int_0^T|t-\lambda^\star|\,|\cos\vartheta_0(t)|\,dt.
\end{aligned}
\label{eq:frequency_NLTR}
\end{equation}
Its minimizer $\lambda^\star$ is the weighted median of $t$ with weight $|\cos\vartheta_0(t)|$ and the optimal HL contribution is
\begin{equation}
J_{\omega|\phi}^{\mathrm{HL}}(T)=D^2(T).
\label{eq:frequency_HL}
\end{equation}
Although the target and nuisance directions coincide instantaneously, $D(T)$ is nonzero because a single $\lambda$ must describe the nuisance ambiguity over the entire evolution.

We now construct the protocol attaining this bound. Introduce a noiseless ancilla and encode the sensor in
$\ket{0_L}=\ket{0}_S\ket{0}_A$ and
$\ket{1_L}=\ket{1}_S\ket{1}_A$.
Continuous QEC corrects the bit-flip errors while retaining the logical Hamiltonian
$H_L(t)=B\sin(\omega t+\phi)Z_L(0)$.
We then apply logical $X_L$ pulses to implement the modulation 
\begin{equation}
y_\star(t)=\operatorname{sgn}
\bigl[(t-\lambda^\star)\cos\vartheta_0(t)\bigr].
\label{eq:frequency_modulation}
\end{equation}
The resulting logical responses are
\begin{equation}
r_\phi(t)=2B y_\star(t)\cos\vartheta_0(t),\qquad
r_\omega(t)=t r_\phi(t).
\label{eq:frequency_responses}
\end{equation}
The weighted-median condition gives $\int_0^T r_\phi(t)\,dt=0$,
and $\int_0^T r_\omega(t)\,dt=D(T)$. The accumulated logical
generators are therefore
\begin{equation}
S_\phi^L(T)=0,\qquad
S_\omega^L(T)=\frac{D(T)}{2}Z_L(0).
\label{eq:frequency_logical_generators}
\end{equation}
Preparing $\ket{\varphi_0^\star}=(\ket{0_L}+\ket{1_L})/\sqrt2$
and measuring $Y_L$ after compensating the nominal evolution
attains Eq.~\eqref{eq:frequency_HL} at the operating point.

In the long-time regime $\omega_0T\gg1$, $|\cos\vartheta_0(t)|$ averages to $2/\pi$, yielding
\begin{equation}
\begin{aligned}
\lambda^\star\simeq\frac{T}{2},\qquad
D(T)\simeq\frac{BT^2}{\pi},\qquad
J_{\omega|\phi}^{\mathrm{HL}}(T)\simeq\frac{B^2T^4}{\pi^2}.
\end{aligned}
\label{eq:frequency_long_time}
\end{equation}
Repeating the protocol $\nu$ times gives 
\begin{equation}
J_{\omega|\phi}^{\mathrm{opt}}\simeq \frac{\nu B^2T^4}{\pi^2}
\end{equation}

It is useful to first compare this result with the idealized case in which $\phi$ is known. The maximal HL contribution under the same noise model is\cite{hu2026optimal}
\begin{equation}
\begin{aligned}
J_{\omega,\mathrm{known}}^{\mathrm{HL,tot}}(T)
&=4\nu B^2\left[\int_0^T t|\cos\vartheta_0(t)|\,dt\right]^2\\
&\simeq\frac{4\nu B^2T^4}{\pi^2},
\end{aligned}
\label{eq:frequency_known_phase}
\end{equation}
where the asymptotic expression holds for $\omega_0T\gg1$. This bound is attained by the same code with $y_0(t)=\operatorname{sgn}[\cos\vartheta_0(t)]$. Thus, an unknown phase reduces the asymptotic QFI by a factor of four but does not alter the $T^4$ scaling.

The importance of treating the nuisance response globally becomes clear if, instead, $\phi$ is treated as an additional error to be corrected locally. The corresponding instantaneous enlarged span is
\begin{equation}
\mathfrak S_{\mathrm n}(t)
=\mathfrak S+\operatorname{span}_{\mathbb R}
\{\partial_\phi H(t)\}.
\label{eq:frequency_enlarged_span}
\end{equation}
Since $\partial_\omega H(t)=t\partial_\phi H(t)$, the target
response lies entirely in $\mathfrak S_{\mathrm n}(t)$.
The INLTR for this enlarged span is therefore
\begin{equation}
\widetilde d_t
=\min_{G\in\mathfrak S_{\mathrm n}(t)}
|\partial_\omega H(t)-G|_s
=\min_\lambda d_t(\lambda)=0,
\label{eq:frequency_enlarged_INLTR}
\end{equation}
and its GNLTR vanishes,
\begin{equation}
\widetilde D(T)=\int_0^T\widetilde d_t\,dt=0.
\label{eq:frequency_local_global}
\end{equation}
Indeed, any code projector $P(t)$ correcting this enlarged span satisfies
\begin{equation}
\begin{aligned}
P(t)\partial_\phi H(t)P(t)&=c_\phi(t)P(t),\\
P(t)\partial_\omega H(t)P(t)&=t c_\phi(t)P(t).
\end{aligned}
\label{eq:frequency_signal_erased}
\end{equation}
so that both responses are proportional to the logical identity. The frequency QFI consequently vanishes,
\begin{equation}
\widetilde J_{\omega|\phi}(T)=\widetilde D^2(T)=0.
\label{eq:frequency_erased_QFI}
\end{equation}
Local removal of the nuisance response therefore erases the target signal itself, whereas our protocol preserves both responses locally and separates them only after accumulation. 

For a second benchmark, consider a protocol designed to maximize the frequency QFI without explicitly accounting for the unknown phase. After correcting the bit-flip noise, the optimal frequency modulation is  $y_0(t)=\operatorname{sgn}[\cos\vartheta_0(t)]$. For a single interrogation time $T$, the resulting QFIM for $(\omega,\phi)$ is
\begin{equation}
\boldsymbol J_{\mathrm{conv}}(T)=4B^2
\begin{pmatrix}
A_1(T)^2&A_1(T)A_0(T)\\
A_1(T)A_0(T)&A_0(T)^2
\end{pmatrix},
\label{eq:frequency_nonrobust}
\end{equation}
where $A_0(T)=\int_0^T|\cos\vartheta_0(t)|\,dt$ and
$A_1(T)=\int_0^T t|\cos\vartheta_0(t)|\,dt$.
The same phase-compensated logical measurement attains this matrix.
Its rank is one, so $J_{\omega|\phi}^{\mathrm{conv}}(T)=0$:
frequency and phase cannot be estimated separately at a single
duration. The conventional protocol therefore combines measurements
at two durations with different ratios $A_1/A_0$.

For a fair comparison, fix the same total interrogation-time resource $\nu T$ and allocate a fraction $s$ to duration $\tau_1$ and the remainder to $\tau_2$. In the large-$\nu$ limit, the corresponding numbers of repetitions are $s\nu T/\tau_1$ and $(1-s)\nu T/\tau_2$, giving
\begin{equation}
\boldsymbol J_{\mathrm{conv}}^{\mathrm{tot}}
=\frac{s\nu T}{\tau_1}\boldsymbol J_{\mathrm{conv}}(\tau_1)
 +\frac{(1-s)\nu T}{\tau_2}\boldsymbol J_{\mathrm{conv}}(\tau_2).
\label{eq:frequency_two_duration_matrix}
\end{equation}
The effective QFI is
\begin{equation}
\begin{aligned}
J_{\omega|\phi}^{\mathrm{conv,tot}}
={}[\boldsymbol J_{\mathrm{conv}}^{\mathrm{tot}}]_{\omega}-\frac{[\boldsymbol J_{\mathrm{conv}}^{\mathrm{tot}}]_{\omega\phi}^{2}}
{[\boldsymbol J_{\mathrm{conv}}^{\mathrm{tot}}]_{\phi}}.
\end{aligned}
\label{eq:frequency_two_duration_optimization}
\end{equation}
For $\omega_0\tau\gg1$,
\begin{equation}
A_0(\tau)\simeq\frac{2\tau}{\pi},\qquad
A_1(\tau)\simeq\frac{\tau^2}{\pi}.
\label{eq:frequency_moments}
\end{equation}
Substituting these expressions into
Eq.~\eqref{eq:frequency_two_duration_optimization} and optimizing
over $0<s<1$ and $0<\tau_1,\tau_2\leq T$ gives
\begin{equation}
\tau_1=T,\qquad \tau_2=\frac{T}{4},\qquad s=\frac13,
\label{eq:frequency_two_duration_design}
\end{equation}
The optimized effective QFI is then
\begin{equation}
J_{\omega|\phi}^{\mathrm{conv,tot}}
\simeq\frac{\nu B^2T^4}{4\pi^2}.
\label{eq:frequency_comparison}
\end{equation}
At the same total interrogation time, this strategy achieves only one
quarter of the effective QFI of our protocol.
\clearpage
\onecolumngrid

\begin{center}
    \mbox{\Large \textbf{Supplemental Material}}

\end{center}

\section{Effective quantum Fisher information in presence of nuisance parameters}

We derive the variational representation of the effective quantum Fisher information (QFI) used in the main text. Consider the estimation of a target parameter $x_1$ in the presence of unknown nuisance parameters $\boldsymbol{x}_{\mathrm n}=(x_2,\ldots,x_N)^T$. For a fixed sensing protocol, the quantum Fisher information matrix (QFIM) can be written in block form as
\begin{equation}
\boldsymbol J_{\boldsymbol{x}}(T)
=
\begin{pmatrix}
J_{x_1}(T) & \boldsymbol J_{x_1x_\mathrm n}(T)\\
\boldsymbol J_{x_\mathrm nx_1}(T) &
\boldsymbol J_{x_{\mathrm n}}(T)
\end{pmatrix}.
\end{equation}
If the nuisance parameters are known, the precision for estimating $x_1$ is governed by $J_{x_1}(T)$. When they are unknown, however, correlations between the target and nuisance responses reduce the information available for estimating $x_1$. For a nonsingular QFIM, the quantum Cram\'er--Rao bound (QCRB) gives
\begin{equation}
\operatorname{Var}(\hat{x}_1)
\ge
\frac{1}{\nu}
[\boldsymbol J_{\boldsymbol{x}}^{-1}(T)]_{11}
=
\frac{1}{\nu J_{x_1|\boldsymbol{x}_{\mathrm n}}(T)},
\end{equation}
where
\begin{equation}
J_{x_1|\boldsymbol{x}_{\mathrm n}}(T)
=
J_{x_1}(T)
-
\boldsymbol J_{x_1x_\mathrm n}(T)
\boldsymbol J_{x_\mathrm{n}}^{-1}(T)
\boldsymbol J_{x_\mathrm n x_1}(T)
\label{eq:supp_schur}
\end{equation}
is the Schur complement of the nuisance block. The second term quantifies the loss of target information caused by parameter variations that cannot be distinguished from nuisance responses.

We now derive an equivalent variational representation of the effective QFI. For $\boldsymbol{\lambda}\in\mathbb R^{N-1}$, consider the quadratic form
\begin{align}
&
\begin{pmatrix}
1 & -\boldsymbol{\lambda}^T
\end{pmatrix}
\boldsymbol J_{\boldsymbol{x}}(T)
\begin{pmatrix}
1\\
-\boldsymbol{\lambda}
\end{pmatrix} =
J_{x_1}(T)
-2\boldsymbol{\lambda}^T\boldsymbol J_{x_\mathrm nx_1}(T)
+\boldsymbol{\lambda}^T
\boldsymbol J_{x_\mathrm{n}}(T)\boldsymbol{\lambda}.
\label{eq:supp_qfi_variation}
\end{align}
For a nonsingular nuisance block, the minimum is attained at
\begin{equation}
\boldsymbol{\lambda}^\star
=
\boldsymbol J_{x_\mathrm{n}}^{-1}(T)
\boldsymbol J_{x_\mathrm nx_1}(T).
\end{equation}
Substituting this result into Eq.~\eqref{eq:supp_qfi_variation}, we obtain
\begin{align}
J_{x_1|\boldsymbol{x}_{\mathrm n}}(T)
&=
\min_{\boldsymbol{\lambda}}
\begin{pmatrix}
1 & -\boldsymbol{\lambda}^T
\end{pmatrix}
\boldsymbol J_{\boldsymbol{x}}(T)
\begin{pmatrix}
1\\
-\boldsymbol{\lambda}
\end{pmatrix}
\nonumber\\
&=
J_{x_1}(T)
-\boldsymbol J_{x_1x_\mathrm n}(T)
\boldsymbol J_{x_\mathrm{n}}^{-1}(T)
\boldsymbol J_{x_\mathrm nx_1}(T),
\label{eq:supp_qfi_variational}
\end{align}
which recovers the Schur complement in Eq.~\eqref{eq:supp_schur}.

When the nuisance block is singular, the effective QFI is given by the generalized Schur complement,
\begin{equation}
J_{x_1|\boldsymbol{x}_{\mathrm n}}(T)
=
J_{x_1}(T)
-
\boldsymbol J_{x_1x_\mathrm n}(T)
\boldsymbol J_{x_\mathrm{n}}^{+}(T)
\boldsymbol J_{x_\mathrm nx_1}(T),
\end{equation}
where $+$ denotes the Moore--Penrose pseudoinverse. Since the QFIM is positive semidefinite, the minimization in Eq.~\eqref{eq:supp_qfi_variation} remains valid, with a minimizer given by
$\boldsymbol{\lambda}^\star
=\boldsymbol J_{x_\mathrm{n}}^{+}(T)
\boldsymbol J_{x_\mathrm nx_1}(T)$.
In particular, if $\boldsymbol J_{x_\mathrm{n}}(T)=0$, the cross block also vanishes, and the effective QFI reduces to $J_{x_1}(T)$.

Combining both cases, we obtain
\begin{equation}
J_{x_1|\boldsymbol{x}_{\mathrm n}}(T)
=
\min_{\boldsymbol{\lambda}\in\mathbb R^{N-1}}
\begin{pmatrix}
1 & -\boldsymbol{\lambda}^{T}
\end{pmatrix}
\boldsymbol J_{\boldsymbol{x}}(T)
\begin{pmatrix}
1\\
-\boldsymbol{\lambda}
\end{pmatrix}.
\label{eq:supp_effective_qfi}
\end{equation}

\section{Nuisance-Aware Continuous-Time Channel-Extension Bound}

We derive the continuous-time channel-extension bound on the
effective QFI used in the main text. For an arbitrary admissible
sensing protocol $\mathcal P$, the variational representation of
the nuisance-aware QFI gives
\begin{equation}
J_{x_1|\boldsymbol{x}_{\mathrm n}}^{\mathcal P}(T)
=
\min_{\boldsymbol{\lambda}\in\mathbb R^{N-1}}
J_{\boldsymbol{\lambda}}^{\mathcal P}(T),
\end{equation}
where
\begin{equation}
J_{\boldsymbol{\lambda}}^{\mathcal P}(T)
=
\begin{pmatrix}
1 & -\boldsymbol{\lambda}^T
\end{pmatrix}
\boldsymbol J_{\boldsymbol{x}}^{\mathcal P}(T)
\begin{pmatrix}
1\\
-\boldsymbol{\lambda}
\end{pmatrix}
\end{equation}
is the QFI along the parameter direction
$\partial_{\boldsymbol{\lambda}}
=\partial_{x_1}-\sum_{j=2}^N\lambda_j\partial_{x_j}$.
Optimizing over all admissible protocols and applying the
minimax inequality yields
\begin{align}
\sup_{\mathcal P}
J_{x_1|\boldsymbol{x}_{\mathrm n}}^{\mathcal P}(T)
&=
\sup_{\mathcal P}
\min_{\boldsymbol{\lambda}}
J_{\boldsymbol{\lambda}}^{\mathcal P}(T)\le
\min_{\boldsymbol{\lambda}}
\sup_{\mathcal P}
J_{\boldsymbol{\lambda}}^{\mathcal P}(T).
\label{eq:supp_minimax}
\end{align}
Thus, for each fixed $\boldsymbol{\lambda}$, it suffices to
bound the single-parameter channel QFI along
$\partial_{\boldsymbol{\lambda}}$.

We apply the continuous-time channel-extension method to the
infinitesimal evolution from $t$ to $t+dt$. For a time-dependent
Hermitian Kraus gauge $h(t)$, let
$\alpha_{\boldsymbol{\lambda}}^h(t)$ and
$\beta_{\boldsymbol{\lambda}}^h(t)$ be the operators defined in
the main text. The infinitesimal channel-extension bound gives
\cite{zhou2018achieving,zhou2021asymptotic}
\begin{equation}
\frac{d}{dt}J_{\boldsymbol{\lambda}}^{\mathcal P}(t)
\le
4\left[
\left\|\alpha_{\boldsymbol{\lambda}}^h(t)\right\|
+
\left\|\beta_{\boldsymbol{\lambda}}^h(t)\right\|
\sqrt{J_{\boldsymbol{\lambda}}^{\mathcal P}(t)}
\right].
\label{eq:supp_differential_ce}
\end{equation}
This bound holds for arbitrary parameter-independent probe
preparations, coherent controls, and adaptive operations.
Since the initial probe is parameter independent,
$J_{\boldsymbol{\lambda}}^{\mathcal P}(0)=0$.
We define
\begin{equation}
A_{\boldsymbol{\lambda}}^h(t)
=
\int_0^t
\left\|\alpha_{\boldsymbol{\lambda}}^h(s)\right\|ds,
\qquad
B_{\boldsymbol{\lambda}}^h(t)
=
\int_0^t
\left\|\beta_{\boldsymbol{\lambda}}^h(s)\right\|ds.
\end{equation}

To integrate Eq.~\eqref{eq:supp_differential_ce}, we consider
the function
\begin{equation}
F(t)
=
4\left[
\sqrt{A_{\boldsymbol{\lambda}}^h(t)}
+
B_{\boldsymbol{\lambda}}^h(t)
\right]^2.
\end{equation}
For $A_{\boldsymbol{\lambda}}^h(t)>0$, its time derivative
satisfies
\begin{align}
\frac{dF(t)}{dt}
&=
4\left\|\alpha_{\boldsymbol{\lambda}}^h(t)\right\|
\left[
1+
\frac{B_{\boldsymbol{\lambda}}^h(t)}
{\sqrt{A_{\boldsymbol{\lambda}}^h(t)}}
\right]
+
4\left\|\beta_{\boldsymbol{\lambda}}^h(t)\right\|
\sqrt{F(t)}
\nonumber\\
&\ge
4\left[
\left\|\alpha_{\boldsymbol{\lambda}}^h(t)\right\|
+
\left\|\beta_{\boldsymbol{\lambda}}^h(t)\right\|
\sqrt{F(t)}
\right].
\end{align}
Equation~\eqref{eq:supp_differential_ce} shows that whenever
the QFI lies below $F(t)$, it also grows no faster than $F(t)$, giving
\begin{equation}
J_{\boldsymbol{\lambda}}^{\mathcal P}(T)
\le
4\left[
\sqrt{A_{\boldsymbol{\lambda}}^h(T)}
+
B_{\boldsymbol{\lambda}}^h(T)
\right]^2.
\label{eq:supp_integrated_ce}
\end{equation}

Since this bound holds for every protocol and gauge,
minimizing over $h$ and applying Eq.~\eqref{eq:supp_minimax}
gives
\begin{equation}
\sup_{\mathcal P}
J_{x_1|\boldsymbol{x}_{\mathrm n}}^{\mathcal P}(T)
\le
4\min_{\boldsymbol{\lambda},h}
\left[
\sqrt{A_{\boldsymbol{\lambda}}^h(T)}
+
B_{\boldsymbol{\lambda}}^h(T)
\right]^2.
\label{eq:supp_nuisance_ce}
\end{equation}

\section{Kraus-Gauge Optimization and Lindblad-Span Distance}

We now evaluate the Kraus-gauge optimization entering the HL contribution and show that it reduces to the distance between the effective Hamiltonian derivative and the instantaneous Lindblad span. For the infinitesimal evolution from $t$ to $t+dt$, the Kraus operators can be written as
\begin{align}
K_0(t,dt)
&=
I-\left[
iH_{\rm tot}(\boldsymbol{x},t)
+\frac{1}{2}\sum_k L_k^\dagger(t)L_k(t)
\right]dt+O(dt^2),
\nonumber\\
K_k(t,dt)
&=
\sqrt{dt}\,L_k(t)+O(dt^{3/2}).
\label{eq:supp_kraus_expansion}
\end{align}
Since the control Hamiltonian and Lindblad operators are independent of the unknown parameters,
\begin{equation}
\partial_{\boldsymbol{\lambda}}K_0
=
-i\partial_{\boldsymbol{\lambda}}H(\boldsymbol{x},t)\,dt
+O(dt^2),
\qquad
\partial_{\boldsymbol{\lambda}}K_k
=
O(dt^{3/2}).
\end{equation}
In the canonical Kraus representation, this gives
$\beta_{\boldsymbol{\lambda}}(t)
=-\partial_{\boldsymbol{\lambda}}H(\boldsymbol{x},t)$.

An equivalent Kraus representation is obtained through a unitary transformation on the Kraus indices,
$\widetilde K_a=\sum_b u_{ab}K_b$.
At the operating point, we may choose $u=I$ and write
$\partial_{\boldsymbol{\lambda}}u=ih$, where $h=h^\dagger$.
The corresponding gauge transformation gives
\begin{equation}
\beta_{\boldsymbol{\lambda}}^h(t)
=
-\partial_{\boldsymbol{\lambda}}H(\boldsymbol{x},t)
-
\lim_{dt\to0}
\frac{1}{dt}
\sum_{a,b}h_{ab}K_a^\dagger K_b.
\label{eq:supp_beta_gauge}
\end{equation}
We take $h_{00}=O(dt)$, $h_{0k}=O(\sqrt{dt})$, and $h_{jk}=O(1)$ for $j,k\geq 1$ so that each block of the Kraus gauge contributes at order $dt$.
Substituting Eq.~\eqref{eq:supp_kraus_expansion} into Eq.~\eqref{eq:supp_beta_gauge}, the gauge-dependent term spans precisely the Hermitian Lindblad span
$\mathfrak S(t)$ generated by
$\{I,L_k(t),L_k^\dagger(t),L_j^\dagger(t)L_k(t)\}_{j,k}$.
Conversely, every Hermitian operator in $\mathfrak S(t)$ can be generated by an appropriate Kraus gauge. Therefore,
\begin{equation}
\min_{h(t)}
\left\|\beta_{\boldsymbol{\lambda}}^h(t)\right\|
=
\min_{G\in\mathfrak S(t)}
\left\|
\partial_{\boldsymbol{\lambda}}H(\boldsymbol{x},t)-G
\right\|.
\label{eq:supp_beta_span}
\end{equation}

Since $I\in\mathfrak S(t)$, an arbitrary identity shift can be absorbed into $G$. For a Hermitian operator, minimizing the operator norm over such shifts gives half its spectral width. Hence
\begin{equation}
\min_{h(t)}
\left\|\beta_{\boldsymbol{\lambda}}^h(t)\right\|
=
\frac{1}{2}
\min_{G\in\mathfrak S(t)}
\left|
\partial_{\boldsymbol{\lambda}}H(\boldsymbol{x},t)-G
\right|_s
=
\frac{1}{2}d_t(\boldsymbol{\lambda}).
\label{eq:supp_beta_distance}
\end{equation}

The Kraus gauge can be optimized independently at each time. Assuming the usual regularity conditions for the time-dependent operators and admissible gauges, the pointwise minimization can be performed under the integral, yielding
\begin{equation}
\min_h B_{\boldsymbol{\lambda}}^h(T)
=
\int_0^T
\min_{h(t)}
\left\|\beta_{\boldsymbol{\lambda}}^h(t)\right\|dt
=
\frac{1}{2}
\int_0^T d_t(\boldsymbol{\lambda})\,dt.
\label{eq:supp_integrated_distance}
\end{equation}
Substituting this relation into the definition of the HL contribution gives
\begin{equation}
J_{x_1|\boldsymbol{x}_{\mathrm n}}^{\mathrm{HL}}(T)
=
\left[
\min_{\boldsymbol{\lambda}\in\mathbb R^{N-1}}
\int_0^T d_t(\boldsymbol{\lambda})\,dt
\right]^2,
\label{eq:supp_optimal_HL}
\end{equation}
which establishes Eq.~\eqref{eq:optimal_HL_precision} of the main text.

\section{Optimal QEC Protocol and Attainability}

We now construct an explicit sensing protocol that attains the optimal HL contribution derived in the main text. Let $\boldsymbol{\lambda}^\star$ minimize $\int_0^T d_t(\boldsymbol{\lambda})\,dt$. We first derive the dual representation of the instantaneous distance. For any Hermitian operator $X$, its spectral width admits the variational representation
\begin{equation}
|X|_s
=
\max_{\rho_0,\rho_1}
\operatorname{Tr}[(\rho_0-\rho_1)X],
\end{equation}
where the maximization is over density operators.Using this identity and convex duality, we rewrite
Eq.~\eqref{eq:instantaneous_distance} as
\begin{align}
d_t(\boldsymbol{\lambda})
&=
\min_{G\in\mathfrak S(t)}
\max_{\rho_0,\rho_1}
\operatorname{Tr}\left[
(\rho_0-\rho_1)
\bigl(\partial_{\boldsymbol{\lambda}}H(\boldsymbol{x},t)-G\bigr)
\right]
\nonumber\\
&=
\max_{\rho_0,\rho_1}
\operatorname{Tr}\left[
(\rho_0-\rho_1)
\partial_{\boldsymbol{\lambda}}H(\boldsymbol{x},t)
\right],
\label{eq:supp_distance_dual}
\end{align}
where the maximization in the second line is restricted
to density operators satisfying
\begin{equation}
\operatorname{Tr}[(\rho_0-\rho_1)G]=0,
\qquad
\forall\,G\in\mathfrak S(t).
\label{eq:supp_dual_constraint}
\end{equation}
After interchanging the optimizations, minimization over $G$
enforces this condition, since otherwise scaling $G$ makes
the objective unbounded below \cite{zhou2018achieving}.

For each time $t$, choose dual-optimal states $\rho_0(t)$ and $\rho_1(t)$ for $\boldsymbol{\lambda}^\star$. We first construct a time-dependent quantum error-correcting code from these states. Let $\ket{\psi_j(t)}$ be purifications of $\rho_j(t)$ using a noiseless ancilla, and introduce orthogonal flag states $\ket{0}_F$ and $\ket{1}_F$. The logical basis is defined by
\begin{equation}
\ket{0_L(t)}
=
\ket{\psi_0(t)}\ket{0}_F,
\qquad
\ket{1_L(t)}
=
\ket{\psi_1(t)}\ket{1}_F.
\label{eq:supp_logical_code}
\end{equation}
We denote the corresponding code projector and logical Pauli operator by
\begin{equation}
P(t)=\ket{0_L(t)}\bra{0_L(t)}
+
\ket{1_L(t)}\bra{1_L(t)},
\qquad
Z_L(t)=
\ket{0_L(t)}\bra{0_L(t)}
-
\ket{1_L(t)}\bra{1_L(t)}.
\end{equation}
The orthogonal flags ensure that all off-diagonal matrix elements of sensor operators between the logical states vanish. Moreover, Eq.~\eqref{eq:supp_dual_constraint} guarantees that the diagonal expectations of every operator in $\mathfrak S(t)$ are identical. Consequently,
\begin{equation}
P(t)GP(t)=c_G(t)P(t),
\qquad
\forall\,G\in\mathfrak S(t).
\label{eq:supp_exact_qec}
\end{equation}
Since $\mathfrak S(t)$ contains the Hermitian and anti-Hermitian components of the Lindblad operators $L_k(t)$ and their products $L_j^\dagger(t)L_k(t)$, Eq.~\eqref{eq:supp_exact_qec} implies the Knill--Laflamme conditions for the infinitesimal noise channel. The Lindblad errors can therefore be removed by continuous recovery in the ideal QEC limit.

Within the logical subspace, the parameter derivatives of the Hamiltonian take the form
\begin{equation}
P(t)\partial_{x_j}H(\boldsymbol{x},t)P(t)
=
\frac{1}{2} a_j(t)P(t)
+
\frac{r_j(t)}{2}Z_L(t),
\label{eq:supp_projected_generator}
\end{equation}
where
\begin{equation}
a_j(t)=
\operatorname{Tr}\left[
\bigl(\rho_0(t)+\rho_1(t)\bigr)
\partial_{x_j}H(\boldsymbol{x},t)
\right].
\end{equation}
The difference between the two diagonal matrix elements gives
\begin{align}
r_j(t)
&=
\bra{0_L(t)}\partial_{x_j}H(\boldsymbol{x},t)\ket{0_L(t)}
-
\bra{1_L(t)}\partial_{x_j}H(\boldsymbol{x},t)\ket{1_L(t)}
\nonumber\\
&=
\operatorname{Tr}\left[
\bigl(\rho_0(t)-\rho_1(t)\bigr)
\partial_{x_j}H(\boldsymbol{x},t)
\right].
\label{eq:supp_parameter_responses}
\end{align}
The first term in Eq.~\eqref{eq:supp_projected_generator} is proportional to the logical identity. The second term determines the relative phase response to $x_j$ retained by the code.

We next construct the optimal probe and coherent control. The initial probe is chosen as the balanced logical superposition
\begin{equation}
\ket{\varphi_0^\star}
=
\frac{\ket{0_L(0)}+\ket{1_L(0)}}{\sqrt2},
\label{eq:supp_optimal_probe}
\end{equation}
which maximizes the variance of $Z_L(0)$. To accumulate the instantaneous responses coherently, we choose a parameter-independent control that transports the logical basis along the time-dependent code,
\begin{equation}
U_{\mathrm{opt}}(t)\ket{j_L(0)}
=
\ket{j_L(t)},
\qquad j=0,1.
\label{eq:supp_optimal_evolution}
\end{equation}
For a piecewise-smooth choice of the logical states, this evolution can be extended to a unitary on the full sensor--ancilla Hilbert space. At a local operating point $\boldsymbol{x}_0$, the corresponding control Hamiltonian is
\begin{equation}
H_C(t)
=
i\dot U_{\mathrm{opt}}(t)U_{\mathrm{opt}}^\dagger(t)
-
H(\boldsymbol{x}_0,t).
\label{eq:supp_optimal_control}
\end{equation}
The operating point is held fixed during each interrogation, so the control is independent of the unknown parameter variations. Together with continuous QEC, this evolution transports the logical basis while preserving the parameter-dependent phase encoding.

Under this control, the logical parameter generators are
\begin{align}
S_{x_j}^L(T)
&=
\int_0^T
U_{\mathrm{opt}}^\dagger(t)
P(t)\partial_{x_j}H(\boldsymbol{x}_0,t)P(t)
U_{\mathrm{opt}}(t)\,dt
\nonumber\\
&=
c_jI_L+
\frac12
\left[\int_0^T r_j(t)\,dt\right]Z_L(0),
\label{eq:supp_accumulated_generators}
\end{align}
where $I_L=P(0)$ and $c_j=\frac{1}{2}\int_0^T a_j(t)\,dt$. The responses are evaluated at $\boldsymbol{x}_0$, and the second equality uses
$U_{\mathrm{opt}}^\dagger(t)Z_L(t)U_{\mathrm{opt}}(t)=Z_L(0)$.

We now show that the dual-optimal states can be chosen so that all accumulated nuisance responses vanish while the target response reaches its optimal value. Substituting
$\partial_{\boldsymbol{\lambda}}H
=
\partial_{x_1}H-\sum_{\ell=2}^N\lambda_\ell\partial_{x_\ell}H$
into Eq.~\eqref{eq:supp_distance_dual} gives
\begin{equation}
d_t(\boldsymbol{\lambda}^\star)
=
r_1(t)
-
\sum_{\ell=2}^N(\lambda^\star)_\ell r_\ell(t).
\label{eq:supp_dual_equality}
\end{equation}
The dual constraints are independent of $\boldsymbol{\lambda}$.
Thus, the states optimal at $\boldsymbol{\lambda}^\star$ remain
feasible at $\boldsymbol{\lambda}^\star+\delta\boldsymbol{\lambda}$.
Evaluating the dual objective on these states and integrating gives
\begin{equation}
\begin{aligned}
\int_0^T
d_t(\boldsymbol{\lambda}^\star+\delta\boldsymbol{\lambda})\,dt
&\geq
\int_0^T d_t(\boldsymbol{\lambda}^\star)\,dt-
\sum_{\ell=2}^N\delta\lambda_\ell
\int_0^T r_\ell(t)\,dt.
\end{aligned}
\end{equation}
The right-hand side is a linear lower bound that becomes exact
at $\delta\boldsymbol{\lambda}=0$. Its slope vector,
$-[\int_0^T r_2(t)\,dt,\ldots,\int_0^T r_N(t)\,dt]^T$,
is a subgradient of the integrated distance.
For integrable Hamiltonian derivatives, the integrated dual
problem is a maximum of linear functions of
$\boldsymbol{\lambda}$, each corresponding to a feasible
time-dependent state pair. At the minimum
$\boldsymbol{\lambda}^\star$, a convex combination of the
slope vectors of the maximizing pairs vanish.
Thus, one common set of weights makes all components
of the accumulated nuisance-response vector zero.

Applying this convex combination to the dual-optimal state pairs,
with the same weights for both states, preserves dual optimality
and Eq.~\eqref{eq:supp_dual_constraint}.
By linearity of the responses, the resulting pair satisfies
all nuisance constraints simultaneously,
\begin{equation}
\int_0^T r_\ell(t)\,dt=0,
\qquad \ell=2,\ldots,N.
\label{eq:supp_vanishing_nuisance}
\end{equation}

Since $\boldsymbol{\lambda}^\star$ is independent of time,
integrating Eq.~\eqref{eq:supp_dual_equality} and using
Eq.~\eqref{eq:supp_vanishing_nuisance} gives
\begin{equation}
\int_0^T r_1(t)\,dt
=
\int_0^T d_t(\boldsymbol{\lambda}^\star)\,dt
=
D(T).
\label{eq:supp_optimal_target_response}
\end{equation}

Substituting Eqs.~\eqref{eq:supp_vanishing_nuisance} and \eqref{eq:supp_optimal_target_response} into Eq.~\eqref{eq:supp_accumulated_generators}, we obtain
\begin{align}
S_{x_1}^L(T)
&=
c_1I_L+
\frac12
\left[
\int_0^T d_t(\boldsymbol{\lambda}^\star)\,dt
\right]Z_L(0),
\nonumber\\
S_{x_\ell}^L(T)
&=
c_\ell I_L,
\qquad \ell=2,\ldots,N.
\label{eq:supp_optimal_generators}
\end{align}
Thus, the nuisance parameters contribute only global phases to first order. The Markovian errors are corrected at each time through Eq.~\eqref{eq:supp_exact_qec}, while the nuisance responses cancel over the full evolution through Eq.~\eqref{eq:supp_vanishing_nuisance}. The instantaneous nuisance responses may remain nonzero.

For the balanced probe in Eq.~\eqref{eq:supp_optimal_probe}, the nuisance-aware QFI becomes
\begin{equation}
\begin{aligned}
J_{x_1|\boldsymbol{x}_{\mathrm n}}^{\mathrm{QEC}}(T)
&=
4\operatorname{Var}\left(S_{x_1}^L(T)\right)\\
&=
\left[
\int_0^T d_t(\boldsymbol{\lambda}^\star)\,dt
\right]^2\\
&=
J_{x_1|\boldsymbol{x}_{\mathrm n}}^{\mathrm{HL}}(T).
\end{aligned}
\label{eq:supp_attained_qfi}
\end{equation}

We now show that the same information is attainable at the measurement level. Consider the projective measurement in the final logical $Y$ basis,
\begin{equation}
\ket{\pm_Y(T)}
=
\frac{\ket{0_L(T)}\pm i\ket{1_L(T)}}{\sqrt2}.
\label{eq:supp_logical_measurement}
\end{equation}
A complete two-outcome POVM on the full Hilbert space is obtained by choosing
\begin{equation}
M_\pm
=
\ket{\pm_Y(T)}\bra{\pm_Y(T)}
+
\frac12\bigl(I-P(T)\bigr),
\end{equation}
which satisfies $M_++M_-=I$.

At the operating point $\boldsymbol{x}_0$, the measurement probabilities satisfy $p_\pm=1/2$. From Eq.~\eqref{eq:supp_optimal_generators}, their local derivatives are
\begin{align}
\left.\partial_{x_1}p_\pm\right|_{\boldsymbol{x}_0}
&=
\pm\frac12
\int_0^T d_t(\boldsymbol{\lambda}^\star)\,dt,
\nonumber\\
\left.\partial_{x_\ell}p_\pm\right|_{\boldsymbol{x}_0}
&=
0,
\qquad \ell=2,\ldots,N.
\label{eq:supp_probability_derivatives}
\end{align}
The resulting classical Fisher information for the target parameter is~\cite{liu2020quantum}
\begin{equation}
\begin{aligned}
F_{x_1}(T)
&=
\sum_\pm
\frac{\left(\partial_{x_1}p_\pm\right)^2}{p_\pm}\\
&=
\left[
\int_0^T d_t(\boldsymbol{\lambda}^\star)\,dt
\right]^2\\
&=
J_{x_1|\boldsymbol{x}_{\mathrm n}}^{\mathrm{HL}}(T).
\end{aligned}
\label{eq:supp_optimal_cfi}
\end{equation}
Since the outcome probabilities are locally insensitive to all nuisance parameters, the measurement extracts the complete target information retained by the optimal QEC protocol.

\section{Functional sensing}

We consider the estimation of a linear function $\theta_1=\boldsymbol{c}^T\boldsymbol{x}$, where $\boldsymbol{c}\neq\boldsymbol{0}$ and $\boldsymbol{x}=(x_1,\ldots,x_N)^T$ denotes the unknown parameters encoded in $H(\boldsymbol{x},t)$. As introduced in the main text, we complete $\theta_1$ into an invertible linear reparametrization
$\boldsymbol{\theta}=(\theta_1,\boldsymbol{\theta}_{\mathrm n}^T)^T=A\boldsymbol{x}$,
where the first row of $A$ is $\boldsymbol{c}^T$ and the remaining rows define the nuisance coordinates. We first show that the effective QFI for estimating $\theta_1$ is independent of the choice of these coordinates.

For any fixed sensing protocol, the QFIM transforms under this reparametrization as
$\boldsymbol J_{\boldsymbol{\theta}}(T)=A^{-T}\boldsymbol J_{\boldsymbol{x}}(T)A^{-1}$.
Using the variational representation derived in Sec.~I, we have
\begin{align}
J_{\theta_1|\boldsymbol{\theta}_{\mathrm n}}(T)
&=
\min_{\boldsymbol{\lambda}}
\begin{pmatrix}
1 & -\boldsymbol{\lambda}^T
\end{pmatrix}
\boldsymbol J_{\boldsymbol{\theta}}(T)
\begin{pmatrix}
1\\
-\boldsymbol{\lambda}
\end{pmatrix}
\nonumber\\
&=
\min_{\boldsymbol{c}^T\boldsymbol{v}=1}
\boldsymbol{v}^T
\boldsymbol J_{\boldsymbol{x}}(T)
\boldsymbol{v},
\label{eq:supp_functional_qfi}
\end{align}
where $\boldsymbol{v}=A^{-1}(1,-\boldsymbol{\lambda}^T)^T$. Since the first row of $A$ is $\boldsymbol{c}^T$, varying $\boldsymbol{\lambda}$ spans all vectors $\boldsymbol{v}$ satisfying $\boldsymbol{c}^T\boldsymbol{v}=1$. A different choice of nuisance coordinates corresponds to a different set of variational coefficients $\boldsymbol{\lambda}'$, which parametrizes the same set of vectors $\boldsymbol{v}$. Consequently, minimizing over $\boldsymbol{\lambda}$ or $\boldsymbol{\lambda}'$ yields the same effective QFI, independently of the choice of nuisance coordinates.

We next show that the optimal sensing framework developed in the main text applies directly to functional sensing under this reparametrization. By the chain rule, the Hamiltonian derivatives satisfy
\begin{equation}
\partial_{\theta_\ell}H
=
\sum_{j=1}^N
(A^{-1})_{j\ell}\partial_{x_j}H,
\qquad \ell=1,\ldots,N.
\end{equation}
Consequently, the effective Hamiltonian derivative becomes
\begin{align}
\partial_{\boldsymbol{\lambda}}^{(\theta)}H
&=
\partial_{\theta_1}H
-
\sum_{\ell=2}^N
\lambda_\ell\partial_{\theta_\ell}H
=
\sum_{j=1}^N
v_j\partial_{x_j}H,
\label{eq:supp_functional_derivative}
\end{align}
where
$\boldsymbol{v}=A^{-1}(1,-\boldsymbol{\lambda}^T)^T$.
Since the Lindblad operators are unchanged under the reparametrization, the corresponding Hermitian Lindblad span $\mathfrak S(t)$ also remains unchanged.

Applying the main result to the reparametrized Hamiltonian, we obtain
\begin{align}
J_{\theta_1|\boldsymbol{\theta}_{\mathrm n}}^{\mathrm{HL}}(T)
&=
\Bigg[
\min_{\boldsymbol{\lambda}}
\int_0^T
\min_{G\in\mathfrak S(t)}
\left|
\partial_{\boldsymbol{\lambda}}^{(\theta)}H-G
\right|_s dt
\Bigg]^2
\nonumber\\
&=
\Bigg[
\min_{\boldsymbol{c}^T\boldsymbol{v}=1}
\int_0^T
\min_{G\in\mathfrak S(t)}
\left|
\sum_{j=1}^N v_j\partial_{x_j}H-G
\right|_s dt
\Bigg]^2.
\label{eq:supp_functional_HL}
\end{align}
The second equality follows from the equivalence of the two minimizations established above. In particular, the optimal HL contribution is independent of the choice of nuisance coordinates.

The attainability of Eq.~\eqref{eq:supp_functional_HL} follows from the same construction as in Sec.~V. The optimal probe, QEC code, coherent control, and measurement are obtained by replacing the effective Hamiltonian derivative with Eq.~\eqref{eq:supp_functional_derivative}. Since the reparametrization leaves the Lindblad dynamics unchanged, the construction applies without further modification and attains the optimal HL contribution in the ideal continuous-QEC regime.

\end{document}